**This article has been published in Appl. Phys. Lett.**

**Please cite this article as:**

Three-dimensional visualization of threading dislocation in GaN by polarized-light microscopy

Yukari Ishikawa[1,*], Kisara Matsumoto[2], Kazuki Ohnishi[2], Yongzhao Yao[1,2]
[*]yukari@jfcc.or.jp

[1]Japan Fine Ceramics Center, 2-4-1, Mutsuno, Atsuta, Nagoya, 456-8587, Japan
[2]Mie University, 1577, Kurimamachiya-cho, Tsu, Mie 514-8507, Japan

We demonstrate high-throughput three-dimensional imaging of threading dislocations in ammonothermal GaN wafers using polarized-light microscopy with collimated LED illumination. Threading dislocations exhibited focal-depth-dependent in-plane shifts, enabling visualization of their three-dimensional inclination behavior over large areas. Synchrotron radiation X-ray topography indicated that dislocations with identical in-plane Burgers vector components tended to exhibit similar trace geometries and inclination directions. The threading dislocations were found to be tilted by approximately 3.3° from the c-axis in directions perpendicular to their Burgers vectors, indicating climb-mediated motion associated with strain relaxation during crystal growth. These results demonstrate a simple nondestructive approach for large-area three-dimensional characterization of dislocations in GaN wafers.

Gallium nitride (GaN) is a wide-bandgap semiconductor with excellent properties such as high breakdown electric field, high carrier mobility, and direct bandgap characteristics, and is widely used in light-emitting and power devices[1–4]. However, crystalline defects remaining in GaN wafers degrade device performance and reliability[5–9]. Therefore, nondestructive characterization of defect distributions over wafer-scale areas is important for both crystal growth optimization and device manufacturing.

Optical microscopy-based techniques are suitable for high-throughput large-area inspection because they can acquire wide-field images in a single shot. Polarized-light microscopy has previously been reported to detect threading dislocations (TDs) in wide-bandgap semiconductors[10–19]. However, previous three-dimensional observation relied on focused-light scanning method[20], which are not suitable for rapid large-area inspection.

In this study, we demonstrate three-dimensional imaging of TDs in GaN using polarized-light microscopy with collimated LED illumination. Furthermore, we analyze the three-dimensional distributions of dislocations and discuss their correlation with strain relaxation behavior.

A commercially available 2-inch basic ammonothermal GaN (0001) wafer[21] with a chemically mechanically polished (CMP) Ga-polar surface was used in this study. After receipt, the N-polar surface was additionally polished by CMP. The wafer thickness was approximately 340 μm. The [0001] axis was tilted by 0.5° toward the $[\bar{1}100]$ direction about the $[11\bar{2}0]$ axis.

The optical system of the polarized-light microscope consisted of a 405 nm LED with a collimating lens, a polarizer with its electric field parallel to $[1\bar{1}00]$, the sample wafer, an objective lens, an analyzer in crossed-Nicol configuration with its electric field parallel to $[11\bar{2}0]$, and a camera. A 20× objective lens (TU Plan Fluor NA 0.45, Nikon) was used. For three-dimensional observation, images were acquired while shifting the objective lens toward the sample in 2.5 μm steps from the surface focus position. Since the focal depth shift inside GaN corresponds to the objective displacement multiplied by the refractive index of GaN ($n = 2.4$), the effective focal shift in the crystal was approximately 6 μm per step. The polarized-light images were processed using FIJI[22].

Image contrast was enhanced using a fast Fourier transform (FFT) bandpass filter followed by contrast adjustment.

Synchrotron radiation X-ray topography (SR-XRT) measurements were performed at BL-3C of KEK Photon Factory. Reflection images were acquired using six equivalent g-vectors of the $11\bar{2}4$ reflections with an X-ray wavelength of 1.4 Å, an incident angle of 5°, and a Bragg angle of 44°.

Figure 1 shows polarized-light microscopy images acquired around a honeycomb defect[21,23] formed by a hexagonally arranged cluster of dislocation bundles (DBs) in ammonothermal GaN. Figure 1 shows a series of images obtained at focal depths from 60 to 270 μm below the Ga-polar-surface in 30 μm increments [Figs. 1(a)–1(h)]. Contrasts corresponding to TDs were clearly observed. The diagonal stripe patterns visible in Figs. 1(a) and 1(h) are not associated with the sample defects; they arise from the infrared autofocus laser pattern superimposed on the polarized-light images during long-exposure acquisition.

The six DBs forming the honeycomb defect are indicated by white arrows. The honeycomb structures formed by DBs exhibited an increasing diameter with increasing depth toward the back surface (N-polar-surface) side. Notably, all 14 honeycomb defects identified in the wafer exhibited similar depth-dependent expansion. In addition to the honeycomb defect, many isolated TDs also exhibited in-plane positional shifts depending on the focal depth. For example, the dislocations indicated by arrows 1–3 moved toward the lower left, lower right, and approximately leftward directions, respectively, as the focal position moved toward the back surface. These focal-depth-dependent positional shifts can be more clearly recognized in Supplementary Video.

The in-plane Burgers vector components of the six DBs forming the honeycomb defect were identified by synchrotron radiation X-ray topography (SR-XRT), as the DBs constituting the honeycomb defect were previously confirmed to be threading edge dislocations (TEDs) in Ref. 23. Reflection topographs were acquired using six equivalent ***g***-vectors of the $11\bar{2}4$ reflection family. Previous ray-tracing simulations have shown that the contrast of edge dislocations depends on the angular relationship between the Burgers vector $\boldsymbol{b}$ and the diffraction vector $\boldsymbol{g}$[24–29]. Based on the reported simulation results, Fig. 2(a) schematically summarizes the contrast relationships for the six equivalent $\boldsymbol{g}$ -vectors corresponding to a single Burgers vector direction[30]. White and black circles represent bright and dark dislocation contrasts, respectively. Since dark contrasts appear for $\boldsymbol{g}_5$ and $\boldsymbol{g}_6$, the Burgers vector direction can be uniquely identified.

Figure 2(b) shows SR-XRT images of the honeycomb defect obtained using the six equivalent reflections. The six DBs forming the honeycomb structure were labeled $D_1$–$D_6$ in a counterclockwise order starting from the upper-right bundle. The dislocations forming $D_1$ exhibited identical contrast behavior under the different diffraction conditions and showed dark contrast for ($\boldsymbol{g}_i$) and ($\boldsymbol{g}_j$), indicating that the in-plane Burgers vector component was parallel to ($\boldsymbol{g}_k$). Similar analyses were performed for $D_2$–$D_6$, and the determined Burgers vector directions are summarized in Fig. 2(c) and Table I.

Figure 3(a) shows a projected image generated from 42 polarized-light images acquired over a depth range from 48 to 294 μm below the surface. The total observation area after stitching was approximately ($1.6 \times 1.4$ mm$^2$). Prior to projection, the polarized-light images were processed using the procedure described in the experimental section. The projected image was then generated by calculating the standard deviation of the image intensity along the $z$-direction at each in-plane ($x$,$y$) position. Bright regions correspond to projections of TDs, namely traces generated by in-plane positional shifts of the dislocation contrasts during focal scanning. Figure 3(b) shows the extracted traces. A total of 768 traces were detected, corresponding to a TD density of approximately ($3 \times 10^4$ cm$^{-2}$).

The inset of Fig. 3(b) shows an enlarged image of the honeycomb defect region enclosed by the blue square. The traces corresponding to $D_1$–$D_6$ exhibited characteristic directions: $D_1$ extended approximately rightward with a slight upward component, $D_2$ upward, $D_3$ approximately leftward with a slight upward component, $D_4$ toward the lower left, $D_5$ downward, and $D_6$ toward the lower right. In addition, the upper-half traces ($D_1$–$D_3$) tended to be shorter than the lower-half traces ($D_4$–$D_6$).

A histogram of the trace angles obtained from the 768 detected traces is shown in Fig. 4(a). As shown in the lower inset of Fig. 4, the trace angle was defined as the direction of the line connecting the contrast position at the front surface to that at the back surface, measured counterclockwise from the $[11\bar{2}0]$ direction, which was defined as 0°. The trace angles were concentrated in six angular ranges of approximately 5–20°, 80–100°, 155–180°, 215–240°, 260–270°, and 290–330°. Figure 4(b) plots the trace lengths as a function of trace angle. Traces with angles of 5–20°, 80–100°, and 155–180° were typically 4–16 μm long, whereas those with angles of 215–240°, 260–270°, and 290–330° ranged from 7 to 35 μm in length.

To explain these distributions, a geometrical model was considered, as illustrated in Fig. 5(a). When TDs are tilted by an angle $\theta$ from the ***c***-axis in directions perpendicular to their Burgers vectors, six equivalent traces are expected at 60° intervals, starting at 30°, provided that the microscope optical axis is parallel to the c-axis. The corresponding traces are expected to have identical lengths. However, the experimental observations differed from the expected symmetry. The traces in the 0–180° range were separated by angular intervals larger than 60°, whereas those in the 180–360° range were separated by intervals smaller than 60°. Furthermore, the traces in the 0–180° range were shorter than those in the 180–360° range. This indicates that the c-axis was slightly inclined toward the $[\bar{1}100]$ direction with respect to the optical axis of the microscope. In practice, achieving perfect alignment between the c-axis and the optical axis is difficult because of the intrinsic c-axis misorientation from the wafer surface normal, sample tilt on the holder, and residual misalignment between the holder and the optical axis.

When the c-axis is tilted toward the $[\bar{1}100]$ direction, the projected trace lengths become asymmetric, with traces extending toward the $[1\bar{1}00]$ side appearing longer and those extending toward the $[\bar{1}100]$ side appearing shorter, as observed in Fig. 5(b). To quantitatively evaluate this effect, the inclination of the c-axis relative to the optical axis was parameterized by angles φ and χ along the $[1\bar{1}00]$ and [1120] directions, respectively. Calculations were performed to identify the combination of φ, χ, and θ that best reproduced the experimentally observed angular distributions and trace lengths (Fig. 4). Among the tested parameter sets, the best agreement with both the angular distributions and trace lengths was obtained for $\varphi = 1.2°$, $\chi \cong 0°$, and $\theta = 3.3°$. The calculated projected lengths were approximately 11 μm for the 10° trace, 12 μm at 90°, 11 μm at 170°, 19 μm at 224°, 18 μm at 270°, and 19 μm at 315° trace. These values are in good agreement with the experimental observations.

The experimentally observed angular distributions were quantitatively reproduced by assuming that the TDs were tilted by approximately 3.3° from the ***c***-axis. The calculated trace lengths also in good agreement with the experimental observations. These results indicate that the TDs in the ammonothermal GaN crystal were tilted approximately 3.3° from the ***c*** -axis in directions perpendicular to their Burgers vectors.

Many isolated TDs outside the honeycomb defect exhibited trace geometries similar to those of $D_1$–$D_6$. Representative dislocations exhibiting these trace geometries were analyzed to examine the relationship between their Burgers vector components and trace directions. Figure 3(c) shows an example SR-XRT image obtained using g =$\bar{1}\bar{1}24$. Assuming that the isolated dislocations were edge dislocations, the Burgers vector directions were determined by comparing their contrast behaviors among the six equivalent reflections. The results are summarized in Table II together with the corresponding trace directions extracted from the polarized-light image stacks. Dislocations exhibiting similar trace geometries were found to possess identical in-plane Burgers vector components and identical in-plane shift directions.

Assuming that these isolated dislocations are edge dislocations, the observed inclination directions were perpendicular to the Burgers vectors, indicating climb-mediated motion rather than glide motion. As the focal position moved from the Ga-polar-surface toward the N-polar-surface, the dislocations shifted outward, corresponding to positive climb motion that reduced the excess half-plane area. The trace geometries observed for the isolated dislocations were similar to those of the corresponding $D_1$–$D_6$ bundles. If these isolated dislocations were assumed to be c+a dislocations, reproducing the observed trace geometries would require a highly specific combination of glide and climb motions while maintaining a constant ratio of their respective components. Such a condition appears less probable, supporting the assumption that the isolated dislocations exhibiting trace geometries similar to those of the corresponding $D_1$–$D_6$ bundles are edge dislocations. The remaining isolated dislocations could not be classified as edge dislocations based on the present analysis and may include c+a dislocations.

When a dislocation advances by a distance $t$ along the $z$-direction, its in-plane displacement is $t \tan \theta$. Therefore, for a dislocation density $\rho$, the total change in the in-plane position of the excess half-planes between the $z = 0$ and $z = t$ planes is expressed as $\rho \, t \tan \theta$. By multiplying this quantity by the lattice spacing along the ***a***-axis, the strain relaxation associated with the dislocation inclination was estimated. Because unpublished analyses by the present authors indicate that TEDs account for approximately 81% of the TDs in the present crystal, the effective strain relaxation associated with dislocation inclination is estimated to be approximately $5.1 \times 10^{-6}$ per millimeter along the growth direction.

Although the detailed crystal growth conditions are unknown, dislocation climb generally occurs under conditions where atomic or vacancy diffusion is thermally activated. Therefore, the observed dislocation inclination is likely associated with strain relaxation induced by thermal stress, temperature inhomogeneity, or impurity concentration gradients during crystal growth. Furthermore, because no preferential inclination direction was observed among the six equivalent directions, the in-plane strain distribution is considered approximately isotropic, whereas a strain gradient likely existed along the ***c***-axis direction.

We demonstrated three-dimensional imaging of TDs in GaN wafers using polarized-light microscopy with collimated LED illumination. The TDs were tilted approximately 3.3° from the c-axis in directions perpendicular to their Burgers vectors, indicating climb-mediated motion. The observed inclination is suggested to contribute to strain relaxation inside the crystal during growth.

## ACKNOWLEDGMENT

This study was partially supported by the Japan Society for the Promotion of Science KAKENHI (Grant No. 23K17356).

## Author Declarations

### Conflict of Interest

The authors have no conflicts to disclose.

### Ethics Approval

This study did not involve human participants or animal subjects and therefore did not require ethics approval.

## AUTHOR CONTRIBUTIONS

**Yukari Ishikawa**: Conceptualization (lead); Data curation (lead); Formal analysis (lead); Investigation (equal); Methodology(lead); Visualization(lead); Writing – original draft (lead); Writing – review & editing (equal).

**Kisara Matsumoto**: Investigation (equal); Formal analysis (equal).

**Kazuki Ohnishi**: Investigation (equal); Resources (lead).

**Yongzhao Yao**; Investigation (equal); Funding acquisition (lead); Writing – review & editing (equal).

## DATA AVAILABILITY

The data that support the findings of this study are available within the article and supplemental material.

Table I. Contrast relationships of the dislocation bundles $D_1$–$D_6$ for $\boldsymbol{g}_{\mathrm{i}}$–$\boldsymbol{g}_{\mathrm{h}}$ reflections and the determined Burgers vectors.

| | $g_i$ $11\bar{2}4$ | $g_j$ $\bar{1}2\bar{1}4$ | $g_k$ $\bar{2}114$ | $g_l$ $\bar{1}\bar{1}24$ | $g_m$ $1\bar{2}14$ | $g_n$ $2\bar{1}\bar{1}4$ | ***b//*** |
|---|---|---|---|---|---|---|---|
| $D_1$ | ● | ● | ○ | ○ | ○ | ○ | $\bar{2}114$ |
| $D_2$ | ○ | ● | ● | ○ | ○ | ○ | $\bar{1}\bar{1}24$ |
| $D_3$ | ○ | ○ | ● | ● | ○ | ○ | $1\bar{2}14$ |
| $D_4$ | ○ | ○ | ○ | ● | ● | ○ | $2\bar{1}\bar{1}4$ |
| $D_5$ | ○ | ○ | ○ | ○ | ● | ● | $11\bar{2}4$ |
| $D_6$ | ● | ○ | ○ | ○ | ○ | ● | $\bar{1}2\bar{1}4$ |

Table II. Determined in-plane $a$-component Burgers vectors of the dislocations 1–6 and the corresponding dislocation shift directions.

| | ***b//*** | | ***Shift direction*** |
|---|---|---|---|
| 1 | $\bar{2}114$ | $D_1$// | Right up |
| 2 | $\bar{1}\bar{1}24$ | $D_2$// | Up |
| 3 | $1\bar{2}14$ | $D_3$// | Right up |
| 4 | $2\bar{1}\bar{1}4$ | $D_4$// | Left down |
| 5 | $11\bar{2}4$ | $D_5$// | Down |
| 6 | $\bar{1}2\bar{1}4$ | $D_6$// | Right down |

Figure caption

Figure 1. Polarized-light microscopy images of the same region acquired at focal depths of (a) 60, (b) 90, (c) 120, (d) 150, (e) 180, (f) 210, (g) 240, and (h) 270 μm.

Figure 2. (a) Schematic illustration of the relationship between the reflection XRT contrast and the $g$-vector for a threading dislocation with Burgers vector $b$. (b) Reflection SR-XRT topographs of the honeycomb defect obtained using equivalent $g$-vectors. (c) Burgers vectors of the dislocation bundles forming the honeycomb defect determined from the SR-XRT contrast analysis.

Figure 3. (a) Projected image generated from 42 polarized-light microscopy images acquired over a depth range from 48 to 294 μm below the surface. (b) Dislocation traces extracted from (a). (c) Reflection X-ray topography image of the same region as in (a) obtained using $g = \bar{1}\bar{1}24$.

Figure 4. (a) Histogram of the inclination angles of the projected dislocation images (traces). (b) Correlation between the inclination angle and length of the projected dislocation images (traces).

Figure 5. Schematic illustration of the relationship between the dislocation inclination directions (upper panels) and the projected dislocation images (traces, lower panels). (a) Case where the $c$-axis is perpendicular to the wafer surface. (b) Case where the $c$-axis is tilted by $\phi$°toward the $[\bar{1}100]$direction.

Supplementary Video 1: which was constructed from polarized-light images acquired through the entire wafer thickness with 6 μm focal-depth intervals.

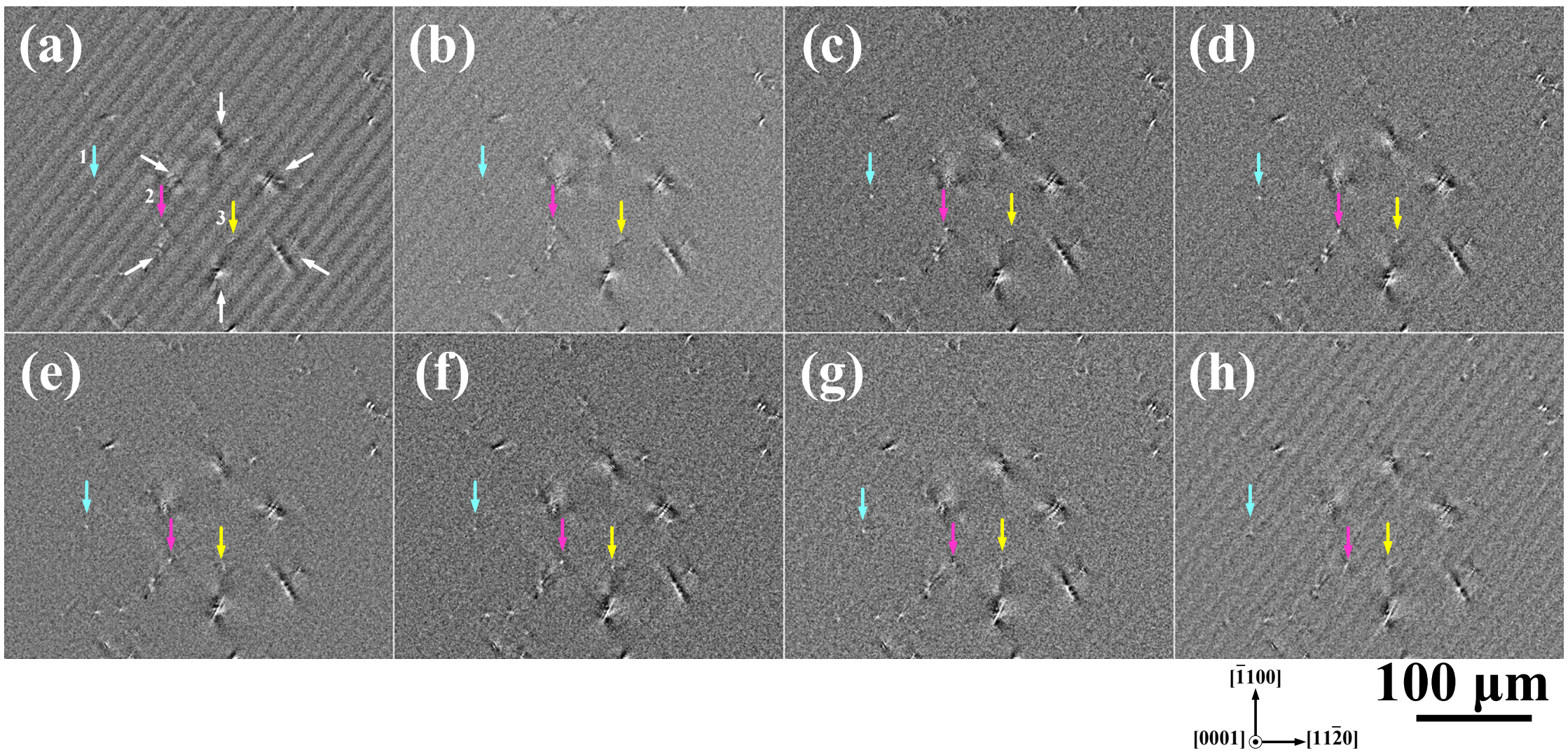

Fig. 1

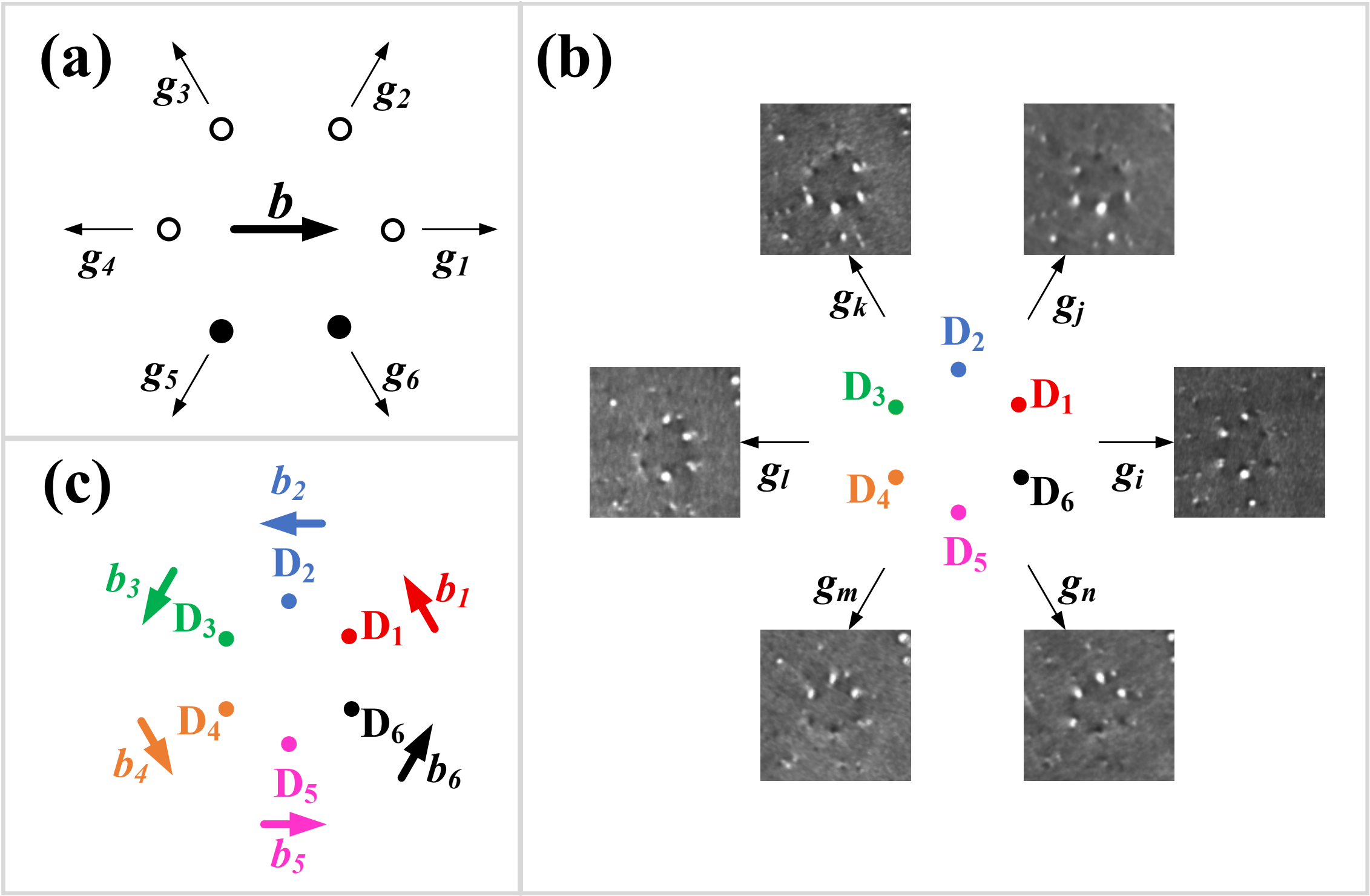

(a)
g3
g2
b
g4
g1
g5
g6
(b)
gk
gj
D2
D3
D1
gl
D4
D6
gi
D5
gm
gn
(c)
b2
D2
b3
D3
D1
b1
D4
D6
b4
b6
D5
b5

Fig. 2

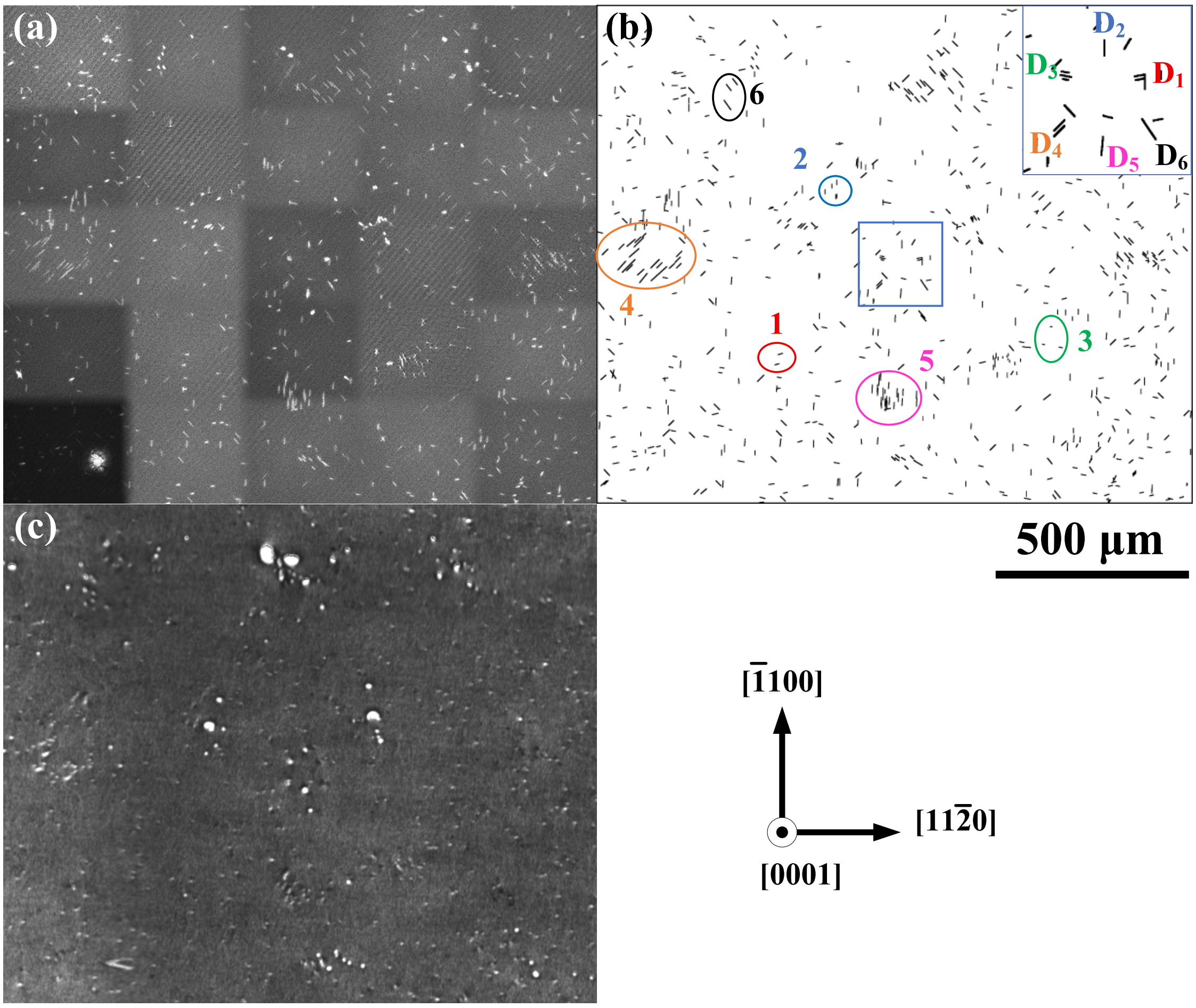

Fig. 3

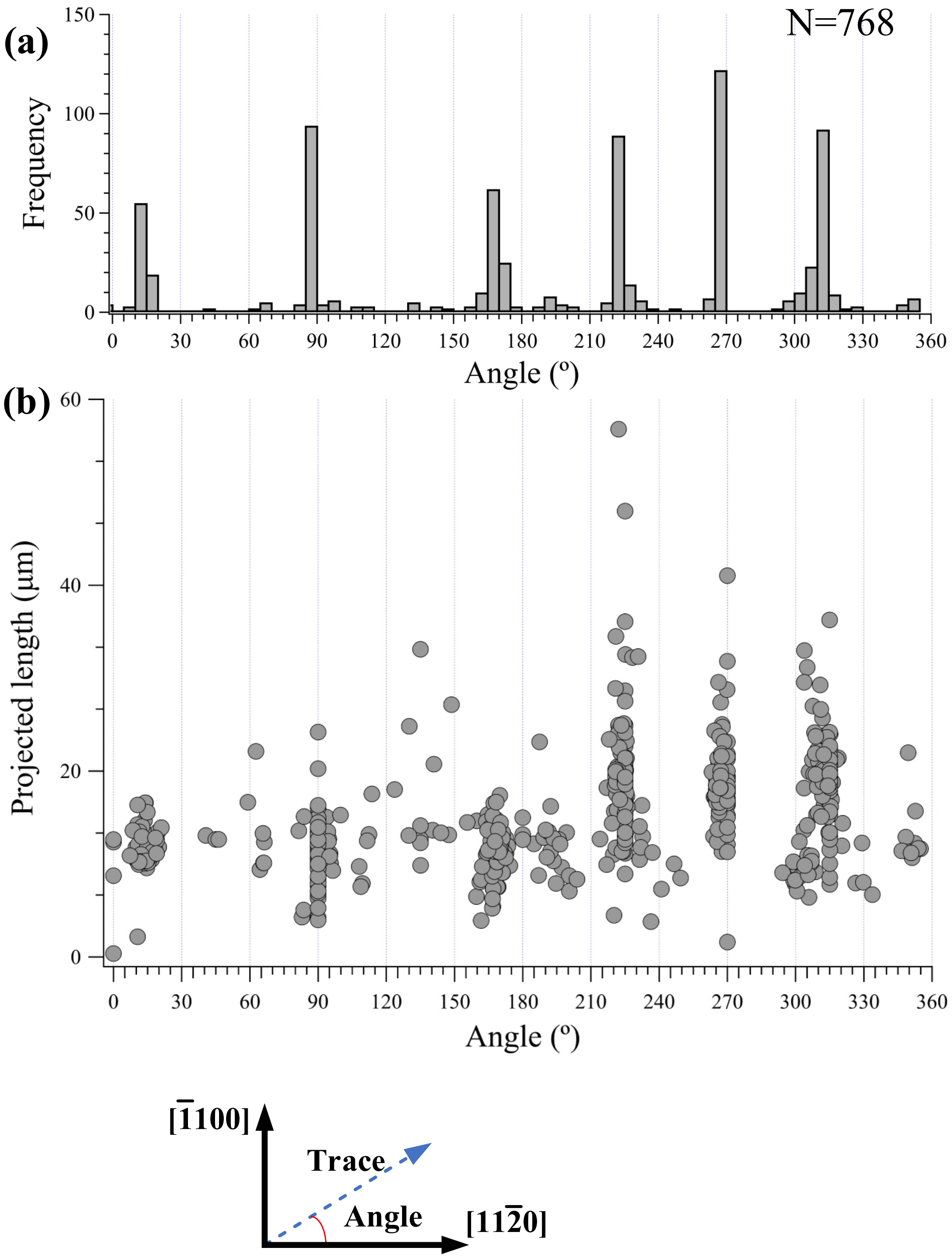
(a)
N=768
Frequency
Angle (º)
(b)
Projected length (μm)
Angle (º)
[1̄100]
Trace
Angle
[112̄0]

Fig. 4

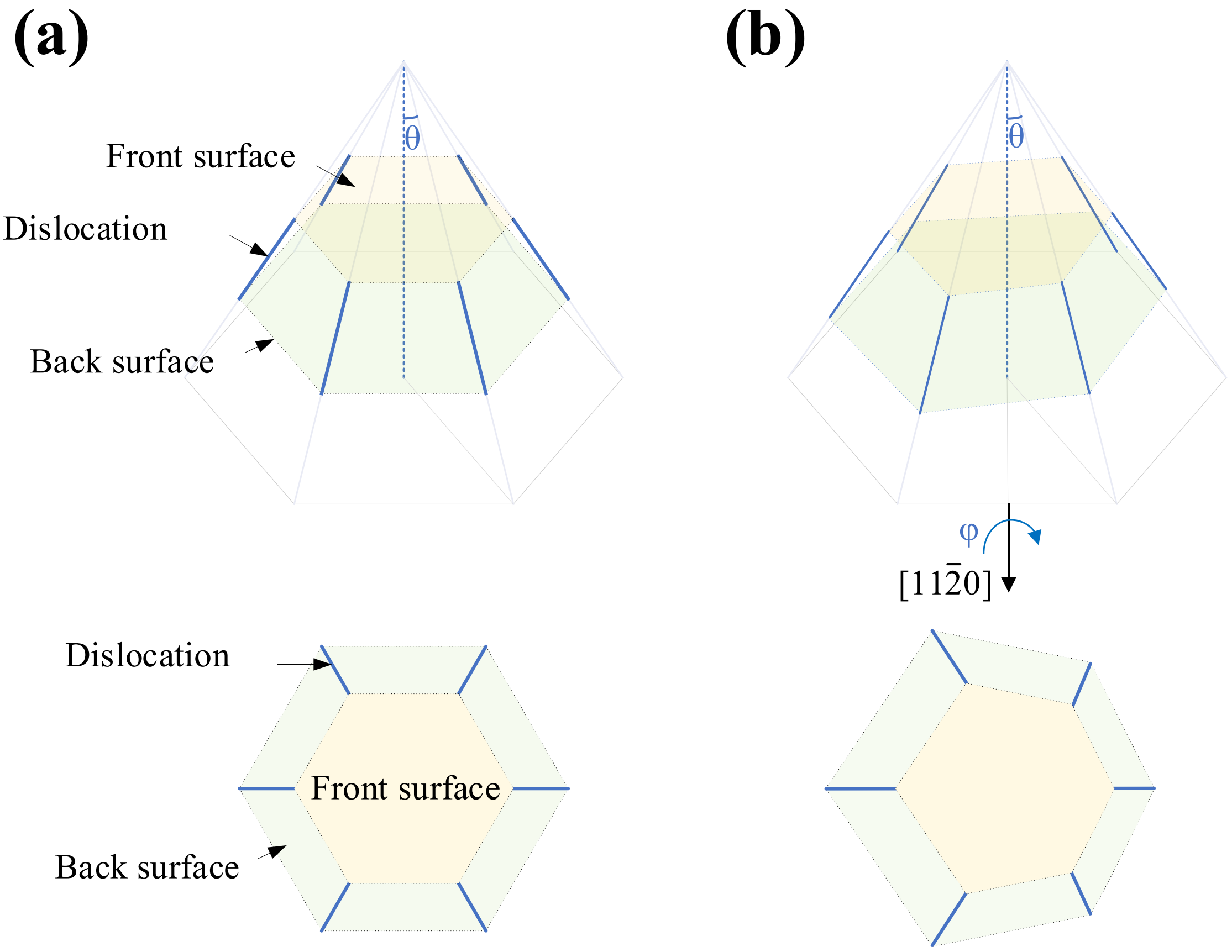
(a)
(b)
Front surface
Dislocation
Back surface
θ
θ
φ
[11$\bar{2}$0]
Dislocation
Front surface
Back surface

Fig. 5